\documentclass{article}

\PassOptionsToPackage{numbers, compress}{natbib}

\usepackage{xcolor}

\usepackage{algorithm}
\usepackage{multicol}
\usepackage{soul}
\usepackage{caption}
\usepackage{comment}
\usepackage[export]{adjustbox}
\definecolor{ourspecialtextcolor}{rgb}{0.528, 0.471, 0.701} %
\usepackage{algorithm}
\usepackage{enumitem}   
\usepackage[noend]{algpseudocode}
\DeclareCaptionSubType*{algorithm}

\algrenewcommand{\algorithmiccomment}[1]{\bgroup\hfill//~#1\egroup}
\algrenewcommand{\Return}{\State\textbf{return}\ }
\algnewcommand{\Save}{\State\textbf{save}\ }
\algnewcommand{\Load}{\State\textbf{load}\ }

\usepackage{ragged2e}
\usepackage{caption}
\usepackage[final]{neurips_2023}
\usepackage{subfig}
\usepackage[utf8]{inputenc} % allow utf-8 input
\usepackage[T1]{fontenc}    % use 8-bit T1 fonts
\usepackage[hyperfootnotes=false]{hyperref}       % hyperlinks
\usepackage{url}            % simple URL typesetting
\usepackage{booktabs}       % professional-quality tables
\usepackage{amsfonts}       % blackboard math symbols
\usepackage{nicefrac}       % compact symbols for 1/2, etc.
\usepackage{microtype}      % microtypography
\usepackage{graphicx}
\usepackage{float}
\usepackage{mathtools}
\usepackage{bbm}
\usepackage{amssymb}
\usepackage{amsmath}
\usepackage{wrapfig}
\usepackage{color, colortbl}
\usepackage{bm}
\DeclareMathOperator*{\argmax}{arg\,max}
\DeclareMathOperator*{\argmin}{arg\,min}

\usepackage{mathtools}

\usepackage{multirow}
\usepackage{bigints}
\usepackage{xspace}

\newcommand{\bR}{\mathbb{R}}
\newcommand{\bPhi}{\bm{\Phi}}

\newcommand{\bx}{\bm{x}}
\newcommand{\by}{\bm{y}}

\newcommand{\bmu}{\bm{\mu}}

\newcommand\blfootnote[1]{%
  \begingroup
  \renewcommand\thefootnote{}\footnote{#1}%
  \addtocounter{footnote}{-1}%
  \endgroup
}

\newcommand{\tgt}{$T$}
\usepackage[textsize=tiny]{todonotes}
\title{On the Out of Distribution Robustness of \\ Foundation Models in Medical Image Segmentation}
\author{Duy M. H. Nguyen$^{*\,1,2,3}$,~Tan N. Pham$^{*
\,3,4}$,~Nghiem T. Diep$^{3}$,~Nghi Phan$^{3}$,\\ \textbf{Quang Pham$^{5}$,~Vinh Tong$^{1,2}$,~Binh T. Nguyen$^{4}$,~Ngan Le$^{6}$,~Nhat Ho$^{7}$,~Pengtao Xie$^{8,9}$},\\ \textbf{Daniel Sonntag$^{3,10}$,~Mathias Niepert$^{1,2}$}
\\\\
$^{1}$University of Stuttgart, $^{2}$IMPRS for Intelligent Systems,\\
$^{3}$German Research Center for Artificial Intelligence, 
$^{4}$University of Science - VNUHCM, \\
$^{5}$Singapore Management University, 
$^{6}$University of Arkansas, 
$^{7}$University of Texas at Austin, \\$^{8}$University of California San Diego, $^{9}$MBZUAI,
$^{10}$Oldenburg University.
}

\begin{document}
\setlength{\abovedisplayskip}{1pt}
\setlength{\belowdisplayskip}{1pt}
\maketitle
\begin{abstract}
Constructing a robust model that can effectively generalize to test samples under distribution shifts remains a significant challenge in the field of medical imaging. The foundational models for vision and language, pre-trained on extensive sets of natural image and text data, have emerged as a promising approach. It showcases impressive learning abilities across different tasks with the need for only a limited amount of annotated samples. While numerous techniques have focused on developing better fine-tuning strategies to adapt these models for specific domains, we instead examine their robustness to
% the robustness of such foundation models to 
domain shifts in the medical image segmentation task. To this end, we compare the generalization performance to unseen domains of various pre-trained models after being fine-tuned on the same in-distribution dataset and show that foundation-based models enjoy better robustness than other architectures. From here, we further developed a new Bayesian uncertainty estimation for frozen models and used them as an indicator to characterize the model's performance on out-of-distribution (OOD) data, proving particularly beneficial for real-world applications. Our experiments not only reveal the limitations of current indicators like \textit{accuracy on the line} or \textit{agreement on the line} commonly used in natural image applications but also emphasize the promise of the introduced Bayesian uncertainty. Specifically, lower uncertainty predictions usually tend to higher out-of-distribution (OOD) performance.
% Our experiments show the shortcomings of existing indicators, such as \textit{accuracy on the line} or \textit{agreement on the line}, commonly employed in natural image applications. Additionally, our findings underscore the potential of the proposed Bayesian uncertainty, indicating that lower uncertainty predictions frequently align with enhanced out-of-distribution (OOD) performance.
% and the promising results of the proposed Bayesian uncertainty, thereby lower uncertainty predictions often correspond to better out-of-distribution (OOD) performance. 

\blfootnote{$^{*}$Co-equal contribution. Corresponding email: \url{Ho_Minh_Duy.Nguyen@dfki.de}}
%Constructing a robust model that can effectively generalize to testing samples under distribution shifts remains a significant challenge in the field of medical imaging. The vision-language foundation model has recently emerged as a promising paradigm, demonstrating impressive zero-shot learning capabilities across various tasks in natural images. However, directly employing these large models in medical domains without fine-tuning often leads to poor performance and necessitates model updates to achieve the expected results.  While numerous approaches have focused on fine-tuning large vision foundation models for specific domains, we instead explore a hypothesis "\textit{Are foundation models robust to domain shifts in medical segmentation tasks after fine-tuning?}" To investigate this, we employ the Segment Anything method (SAM), which has been trained on over 1 billion annotations from natural images, and assess its performance across three realistic domain shifts in medical segmentation tasks. Our evaluation of SAM encompasses diverse settings, including prompt-based segmentations and end-to-end segmentations, with a comparison against popular supervised learning baselines. Moreover, we empirically estimate the distribution shifts between datasets and examine their correlation with the obtained results.
\end{abstract}

\section{Introduction}
\vspace*{-0.05in}
\if0
\textcolor{red}{Duy: Some important factors:
\begin{itemize}
    \item We first conduct systematic experiments to validate large vision model performance under medical segmentation distribution shift settings. The results show that these large vision models are more robust than other pre-trained ImageNet models.
    \item We propose a new approach to predict the model's performance in out-of-distribution for segmentation tasks by estimating uncertainty in predictions. To this end, we leverage the Bayesian identity calibration techniques,  a class of methods that estimate uncertainty for frozen models without requiring training models from scratch - and tailor them specifically to the challenges posed by segmentation problems. It is important to note that while previous studies report a linear correlation between model performance on in-distribution (ID) and out-of-distribution (OOD) or model invariances and OOD, none of them hold in our settings. In reverse, we indicate that uncertainty quantification can be used as surrogate estimations. 
\end{itemize}
}
\fi

Recent years have witnessed tremendous success of foundation models~\citep{bommasani2021opportunities, floridi2020gpt, devlin2018bert}, which have greatly impacted research in many domains, ranging from understanding language~\citep{touvron2023llama}, to vision~\citep{oquab2023dinov2}. Such models are pre-trained on massive datasets and have shown encouraging capabilities in performing many tasks, even when only fine-tuned on a small number of samples ~\citep{kirillov2023segment}.
Since then, foundation models have presented unprecedented opportunities for researchers to explore more challenging and impactful problems.
%With such models, researchers now have immediate access to good feature representations to address more challenging and impactful problems.
Among these, medical image understanding~\citep{willemink2022toward} has been an attractive topic due to its potential impacts on our society. However, because of the intrinsic differences between medical and natural images, directly applying models pre-trained on natural images to the medical domain may lead to unsatisfactory results~\citep{nguyen2022tatl,ma2023segment,shi2023generalist}. Thus, it is imperative to investigate the transferability and robustness of foundation models to unlock their full potential for real-world medical applications.

% Figure~\ref{fig:domain-shift} illustrates several image samples collected from different domains.

\begin{figure}
    \centering
    \includegraphics[width=1.0\textwidth]{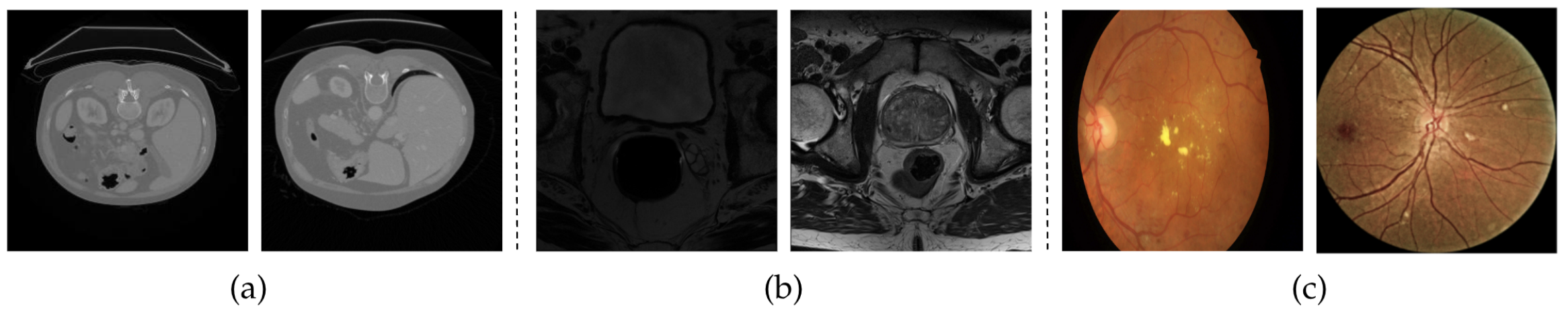}
    \caption{Illustration of domain shift in medical segmentation. From left to right are pairs of liver (a), prostate (b), and diabetic retina images (c) selected from different datasets.} 
    \label{fig:domain-shift}
    \vspace*{-0.2in}
\end{figure}

%such as analyzing medical images~\citep{willemink2022toward}. However, because of the differences in medical and natural images, directly applying such pre-trained models to medical imaging tasks may render unsatisfactory results~\citep{nguyen2022tatl,ma2023segment,shi2023generalist}. 
%Therefore, investigating and improving the transferability of foundation models in medical domains has been a crucial research direction in fully realizing the potential of such foundation models.
In this work, we investigate the ability to generalize to unseen distributions of various deep learning models, especially large foundation models, in the medical image segmentation task. To this end, we first consider several popular architectures based on UNet and MedSAM. UNet and its derivatives, including UNet++ \citep{zhou2018unet++} or TransUNet \citep{chen2021transunet}, have conventionally served as prevalent approaches in medical image segmentation. In contrast, MedSAM~\citep{huang2023segment} is a recent method that focuses on \textit{fine-tuning} the Segment Anything Model (SAM)~\citep{kirillov2023segment}), which is one of the best publicly available models for generic segmentation~\citep{minaee2021image}, on a specific medical dataset.
% While UNet~\citep{ronneberger2015u} and its variants~\citep{zhou2018unet++, chen2021transunet} have been a common approach to medical image segmentation, MedSAM~\citep{huang2023segment} is a recently introduced method to \textit{finetune} the Segment Anything Model (SAM)~\citep{kirillov2023segment}), which is one of the best publicly available models for generic segmentation~\citep{minaee2021image}, on a specific dataset. 
We comprehensively evaluate such models on various medical image segmentation tasks by training them on the source domain and then performing evaluations on target domains, which come from different distributions as illustrated in Figure~\ref{fig:domain-shift}. Across all datasets, the experiments showed that fine-tuned foundation models such as MedSAM offer better generalization to unseen domains than traditional models pre-trained on ImageNet. These results demonstrate the promising capabilities of foundation models for real-world deployment.

For a reliable real-world deployment, however, one needs to estimate the model's OOD performance without actually training the model on the target domain. For example, consider a collection of pre-trained models that can segment liver images; given a hospital in a different location, one would like to select a model that will yield the best segmentation results for the patients at this location without training all the models in the collection. However, the lack of labeled samples in the target domain~\citep{miller2021accuracy, deng2022strong} necessitates an effective strategy to model the OOD performance using only unlabeled data. To investigate this problem, we first consider several popular indicators that have shown promising results with natural images, such as the in-domain (ID) performance~\citep{miller2021accuracy} or the agreement with another network~\citep{baek2022agreement}. Interestingly, our findings show that none of such indicators are applicable in the medical image segmentation tasks as they do not hold linear correlations to the OOD performance as expected (Figure \ref{fig:id-ood-linear}).
% This motivates us to propose a Bayesian uncertainty estimator tailored to segmentation tasks to seek a more reliable indicator to predict the OOD performance. 
This motivates us to introduce a tailored Bayesian uncertainty estimator designed specifically for segmentation tasks, aiming to provide a more dependable indicator for predicting out-of-distribution (OOD) performance. Our experimental results indicate that higher uncertainty in the model's predictions consistently reflects lower OOD performance. In summary, we shed light on the challenges associated with accurately estimating OOD performance in medical image segmentation tasks, underscore the limitations of conventional indicators applied in natural image contexts, and demonstrate the promising results achieved through the proposed Bayesian uncertainty as a surrogate estimator.

\section{Methodology}
% \vspace{-0.15in}
\subsection{Notations and Settings}
\noindent
% \textbf{Notation}
We denote $D = \{(\bx_i, \by_i)\}_{i=1}^{N} \overset{\text{i.i.d}}{\sim} S$ as a labeled training set of $N$ samples which are independently and identically distributed (i.i.d) sampled from a source domain $S$. Here, $(\bx_{i}, \by_{i})$ represents a pair consisting of an image (flattened to become a vector) $\bx_i \in \mathbb{R}^{m}$ and its label $\by_{i} \in \mathbb{R}^{m}$. We introduce a deep network denoted as $\bm{\Phi}(.;\theta): \bR^{m} \rightarrow \bR^{m}$, parameterized by $\theta$, which maps the images from the set $\bx_{i}$ to the desired outputs $\by_{i}$. The primary objective of our learning process is to train a model $\bPhi$ using the training dataset $D$ where $\theta$ is possibly initialized from foundational models (SAM) or pre-trained models like ImageNet, which are trained on large amounts of natural images. The trained model will then be evaluated for accuracy when applied to samples from a different target domain \tgt. In our setup, the source and target domains, $S$ and $T$, respectively, are chosen from datasets that pertain to the same organ. However, these datasets experience domain shift issues due to variations in scanner devices or acquisition conditions (Figure \ref{fig:domain-shift}).
\begin{figure}[!ht]
    \centering
    \includegraphics[width=1.0\textwidth]{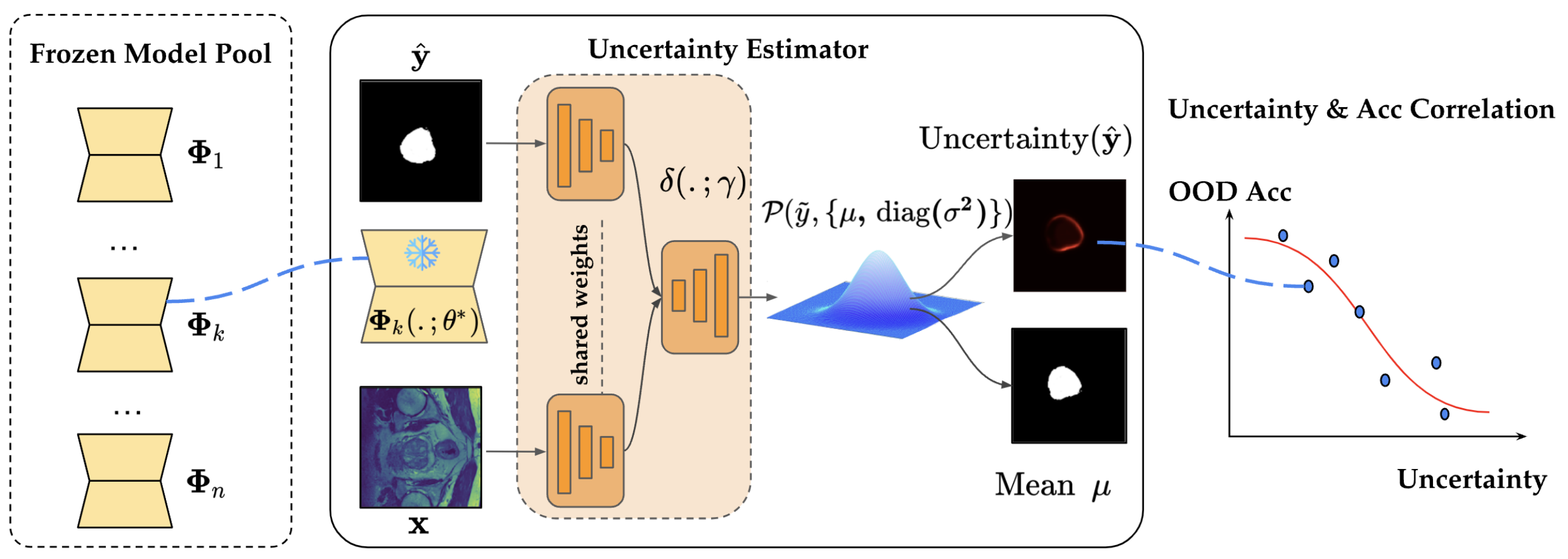}
    \caption{Predicting OOD performance using uncertainties for a pool of frozen models. Each trained network $\bPhi_{k}(.;\theta^{*})$ on a source domain is paired with an independent network $\bm{\delta}(.;\gamma)$ which takes both input images $\bx$ and output $\hat{\by} = \bPhi_{k}(\bx;\theta^{*})$ and produces uncertainty for each prediction $\hat{\by}$ of $\bPhi_{k}(.;\theta^{*})$ in a target domain. The uncertainty values are extracted to estimate a correlation to OOD performance in the target domain for each model.} 
    \vspace{-0.2in}
    \label{fig:our_model}
\end{figure}

\vspace{-0.3in}
\subsection{Prompt-based Segmentation}
\label{subsec:prompt-end-2-end}
We examine the generalization capabilities of SAM for a prompt-based segmentation scenario. We follow the MedSAM method~\cite{ma2023segment} to freeze the image and prompt encoder layers; only mask decoder layers are learned during the fine-tuning phase. The prompt layers use box-based prompts generated by rectangles covering ground-truth regions perturbed with random offsets. We then train it with training data in the single source domain $S$ and take the trained network to predict the test set of the target domain $T$. Implementation details are presented in Section \ref{subsec:sam-architecture} Appendix. 

\subsection{Performance Prediction on Out-of-Distribution under Uncertainty Perspective}
Our research proposes a new perspective to explore the relationship between in-distribution (ID) and OOD performance by quantifying the model's uncertainty in the target domains.
% On Distribution (ID) OOD and in distribution performance by observing model uncertainty.
% Our research pioneers a new perspective by exploring the relationship between OOD model performance and model uncertainty, shedding light on an unexplored aspect of machine learning models. 
% Our findings challenge conventional wisdom, as they reveal that higher uncertainty often corresponds to poorer OOD performance. 
Unlike previous research \citep{miller2021accuracy, deng2022strong, baek2022agreement}, which mainly focused on image classification tasks and proposed indicators to correlate ID and OOD linearly, our settings in medical image segmentation tasks highlight a crucial difference: none of the previously suggested indicators are applicable in our domain, as demonstrated in Figure \ref{fig:id-ood-linear}, where relatively modest Pearson correlation coefficients are observed. This underscores the need for a different approach specific to medical image analysis to capture such complex relationships. Our focus delves into these challenges and provides a proof of concept, indicating that leveraging Bayesian techniques for uncertainty estimation in frozen models can effectively forecast model performance in out-of-distribution scenarios, eliminating the need for labeled samples.
\vspace{-0.1in}
\paragraph{\textbf{Formulation}:}
Given a trained model $\bPhi(,;\theta^{*})$, for an input $\bx_{i} \in T$, our goal is to model a distribution of $\mathcal{P}(\tilde{\by}|\bx_i)$ rather than a point estimate $\hat{\by}_i = \bPhi(\bx_i;\theta^{*})$. While Bayesian deep learning-based approaches \cite{daxberger2021laplace} can provide such uncertainty estimations, they require predefined architectures and have to train models from scratch \cite{godard2017unsupervised,dosovitskiy2015flownet}. However, in our context, accessing complete training data during the pre-trained stage is not feasible, as seen in foundation models, and training costs can be prohibitively high. To tackle this challenge, we turn to the latest post-hoc techniques~\citep{daxberger2021laplace,eschenhagen2021mixtures,upadhyay2022bayescap, rangnekar2023usim, upadhyay2023hypuc, upadhyay2023probvlm} for estimating uncertainty in tasks such as image translation, image enhancement, and depth prediction in self-driving cars. This technique has recently been adapted to active learning research~\citep{rangnekar2023usim} and Vision-Language Models~\citep{upadhyay2023probvlm}. We tailor these methods for medical segmentation, where regions of interest are often small and surrounded by similar tissue structures. We depict our method in Figure \ref{fig:our_model}.
% which is not the case in our settings given the fact that we cannot access fully training data in the pre-trained stage (e.g., in foundation models) or due to the prohibitive training costs. To address this problem, we resort to advanced post-hoc methods [] that have recently been proposed to estimate uncertainty in image translation, image enhancement, or depth prediction in self-driving cars and tailor them properly for medical segmentation where the interest regions are typically small and are dominated by surrounding similar tissue structures. 

In particular, we construct an probabilistic model $\bm{\delta}(.;\gamma)$ to estimate uncertainty for the frozen model $\bPhi(.;\theta)$ by producing an independent multivariate Gaussian distribution $\mathcal{P}\left(\tilde{\by};\{\bm{\mu}_i, \mathrm{diag}(\bm{\sigma}_i^{2})\}\right)$ parameterized by $\{\bm{\mu}_i, \bm{\sigma}_i^2\} = \bm{\delta}(\{\hat{\by}_i, \bx_i\};\gamma)$ for every input-output pair $\{\bx_i, \hat{\by}_i\}$. This distribution conveys information about the possible values of the reconstructed output $\tilde{\by}_i$ and the uncertainty $\bm{\sigma}^2_i$ of the prediction. We seek to optimize the uncertainty estimator to maximize the data likelihood:

\begin{align}
    \gamma^{*}  & = \argmax\limits_{\gamma} \prod_{i=1}^{N}\,\mathcal{P}\left({\by_{i}};\{\bm{\mu}_i, \mathrm{diag}(\bm{\sigma}_i^{-2})\}\right) \\
    & = \argmax\limits_{\gamma} \prod_{i=1}^{N}\,\frac{1}{(2\pi)^{m/2}\left(\prod_{j=1}^{m} \sigma_{ij}^2\right)^{\frac{1}{2}}}\exp\left(-\frac{1}{2}({\by_i} - \bm{\mu}_i)^{\top}\mathrm{diag}(\bm{\sigma}_i^{-2})(\by_i - \bm{\mu}_i)\right) \\
    &= \argmax\limits_{\gamma} \prod_{i=1}^{N}\, \prod_{j=1}^{m} \frac{1}{(2\pi)^{1/2} \sigma_{ij}}\,\exp\left(-\frac{1}{2\sigma_{ij}^2}({\by_{ij}} - \bm{\mu}_{ij})^{2}\right) \label{eq:one-gauss}\\
    &= \argmin\limits_{\gamma} \sum_{i=1}^{N}\, \sum_{j=1}^{m}\,\frac{({\by_{ij}} - \bm{\mu}_{ij})^{2}}{2\sigma_{ij}^2} + \frac{\log{\sigma_{ij}^2}}{2}. \label{eq:loss}
\end{align}

In order to reconstruct output of $\bPhi(,;\theta)$
from $\bm{\delta}(.;\gamma)$, we can extend Eq.\eqref{eq:loss} to:

\begin{align}
     \gamma^{*} = \argmin\limits_{\gamma}\,\sum_{i=1}^{N}\, \sum_{j=1}^{m}\,\frac{({\by_{ij}} - \bm{\mu}_{ij})^{2}}{2\sigma_{ij}^2} + \frac{\log{\sigma_{ij}^2}}{2} + \lambda \|\bm{\mu}_i - \hat{\by}_i\|^2 \label{eq:final-loss}
\end{align}

where $\lambda$ indicates the hyper-parameter controlling the trade-off between maximum likelihood and reconstructed output contributions. After optimizing Eq.\eqref{eq:final-loss} on samples of source domain $S$, we then can compute uncertainty for predictions $\hat{\by}_t = \bPhi(\bx_{t};\theta^{*})$ with $\bx_{t} \in T$ by:

\begin{align}
   \mathrm{Uncertainty}(\hat{\by}_t)  = \bm{\sigma}_t^2 \in \bR^{m}, \; \text{where } \{\bm{\mu}_t,\,\bm{\sigma}_t^{2}\}  = \bm{\delta}(\{\hat{\by}_t, \bx_t\};\gamma^{*})\label{eq:uncertainty_model}. 
\end{align}
\noindent
\paragraph{\textbf{Generalized Uncertainty with Heavy-Tailed Distribution}:}
Note that Eq. \eqref{eq:one-gauss} can be seen as a product of univariate Gaussians $\mathcal{N}(\bm{\mu}_{ij},\sigma_{ij}^2)$ modeling distributions at the per-pixel level. Therefore, one can extend them to generalized Gaussian distributions \cite{dytso2018analytical} $\mathcal{GGD}(\bm{\mu}_{ij},\alpha_{ij},\,\beta_{ij})$ with scale $\alpha_{ij}$ and shape $\beta_{ij}$ coefficients used to model heavy-tailed distribution at pixels which usually occur due to the presence of noises or artifacts~\citep{upadhyay2022bayescap, upadhyay2023hypuc, upadhyay2023probvlm}.
% We then obtain a generalized loss function:
% \begin{align}
%      \gamma^{*} = \argmin\limits_{\gamma}\,\sum_{i=1}^{N}\, \sum_{j=1}^{m}\,\left(\frac{\left|{\by_{ij}} - \bm{\mu}_{ij}\right|}{\alpha_{ij}}\right)^{\beta_{ij}}-\log\,\frac{\beta_{ij}}{\alpha_{ij}} + \log\,\Gamma\left(\beta_{ij}^{-1}\right) + \lambda \|\bm{\mu}_i - \hat{\by}_i\|^2. \label{eq:generalized-loss}
% \end{align}
% where $\Gamma(z) = \bigintsss_{\,0}^{\mathrm{inf}}x^{z-1}\exp(-x)dx,\ \forall z > 0$ is the Gamma function.
The uncertainty for the prediction of $\bx_t \in T$ in this case is computed as:

\begin{align}
   & \{\bm{\mu}_t,\,\bm{\beta}_t, \bm{\alpha}_{t}\}  = \bm{\delta}(\{\hat{\by}_t, \bx_t\};\gamma^{*});\ \bm{\beta}_t = \left[\beta_{t1},...,\beta_{tm}\right]; \bm{\alpha}_{t} = \left[\alpha_{t1},...,\alpha_{tm}\right] \label{eq:alpha_beta_ggd_1}\\
   & \mathrm{Uncertainty}(\hat{\by}_t)  = \left[\frac{\alpha_{t1}^2\Gamma\left(3\beta_{t1}^{-1}\right)}{\Gamma\,\left(\beta_{t1}^{-1}\right)},...,\frac{\alpha_{km}^2\Gamma\left(3\beta_{tm}^{-1}\right)}{\Gamma\,\left(\beta_{tm}^{-1}\right)}\right]^\top \in \bR^{m}. \label{eq:alpha_beta_ggd_2}
\end{align}

where $\Gamma(z) = \bigintsss_{\,0}^{\mathrm{inf}}x^{z-1}\exp(-x)dx,\ \forall z > 0$ is the Gamma function.

\subsection{Comparison to Existing  Models for Post-hoc Uncertainty Quantification}
\vspace{-0.1in}
Our formulation is similar to that of existing approaches~\citep{upadhyay2022bayescap, upadhyay2023hypuc, upadhyay2023probvlm} in that we also estimate the uncertainty of the frozen model $\bPhi(.;\theta^{*})$ by training an auxiliary network $\bm{\delta}(.;\gamma)$ with the same objective. However, instead of conditioning only on the output  $\hat{\by}_{i} = \bPhi(\bx_{i};\theta^{*})$, i.e., $\mathrm{Uncertainty}(\hat{\by}_{i}) \sim \bm{\delta}(\hat{\by}_{i};\gamma^{*})$, we
propose a  model for uncertainty quantification $\bm{\delta}(.;\gamma)$ driven by both $\hat{\by}_{i}$ and the original image input $\bx_{i}$ for the medical segmentation by
 $\mathrm{Uncertainty}(\hat{\by}_{i}) \sim \bm{\delta}(\{\hat{\by}_{i}, \bx_{i}\};\gamma^{*})$. This is motivated by the following observations. 
 
 First, in segmentation settings, the outputs $\hat{\by}_{i}$ produced by the frozen models are simply binary masks and, therefore, there is no surrounding context for $\bm{\delta}(.;\gamma)$ to learn the maximum likelihood conditions as in Eq.\eqref{eq:loss}. In other words, the model tends to reconstruct only the output of frozen models $\bPhi(,;\theta^{*})$ by minimizing $\|\bmu_{i} - \hat{\by}_{i}\|^2$ while tending to over-fit the remaining components in Eq.\eqref{eq:final-loss}. This is essentially different from other methods~\citep{upadhyay2022bayescap, upadhyay2023probvlm, upadhyay2023hypuc} designed for image enhancement, image translation, or depth estimation, wherein outputs $\hat{\by}_i$ are continuous signals and have high mutual information with input images $\bx_{i}$.   

Second, under a probabilistic view, we have:
\begin{equation}
    \mathcal{P}\left(\mathrm{Uncertainty}(\hat{\by}),  \hat{\by}\, |\,\bx \right) = \mathcal{P}\left(\mathrm{Uncertainty}(\hat{\by})\, |\,\hat{\by},\bx \right) \cdot \mathcal{P}\left(\hat{\by}\, |\,\bx \right).
\end{equation}

Both our model $\bm{\delta}(;\gamma) $ and existing methods estimate $\mathcal{P}\left(\hat{\by}|\bx\right)$ by reconstructing the output of the frozen model $\hat{\by} = \bPhi(\bx|\theta^{*})$ trained 
% so as to learn $\mathcal{P}\left(\hat{\by}|\bx\right)$
during the pre-training step. However, the remaining factor $\mathcal{P}\left(\mathrm{Uncertainty}(\hat{\by})\, |\,\hat{\by},\bx \right)$ is learnt by utilizing pairs of $\{\hat{\by},\bx\}$ in $\bm{\delta}(;\gamma) $ while other methods  approximate $\mathcal{P}\left(\mathrm{Uncertainty}(\hat{\by})\, |\,\hat{\by},\bx \right) \approx \mathcal{P}\left(\mathrm{Uncertainty}(\hat{\by})\, |\,\hat{\by}\right)$. This approximation fails, however, if  $\hat{\by}$ does not contain sufficient information about the input $\bx$ as is the case in the segmentation settings.

Finally, instead of computing averaging uncertainty values as prior works,
% in contrast to prior methods that compute average uncertainty values,
we utilize Otsu's parameter-free thresholding algorithm~\cite{otsu1979threshold} on the uncertainty matrix. This identifies pixel groups with the highest uncertainty, allowing us to measure their areas. Our approach, particularly effective for small regions of interest in segmentation tasks, exhibits a stronger correlation to OOD scenarios.

\begin{table}[H]
\centering
\setlength{\tabcolsep}{6pt}
\renewcommand{\arraystretch}{0.7}
\caption{Performance comparison in domain-shift of prostate segmentation. Results are reported in average 3D IoU score with three training times. The BMC and BIDMC datasets are selected as source domains, and the remaining datasets are used as target domains. Arrows $\uparrow$ and $\downarrow$ indicate the increase/drop performance between target and source domains.}
\vspace{2mm}
\resizebox{1.0\textwidth}{!}{
\begin{tabular}{l|l|ccccc|c}
\toprule
 & & 
  \textbf{UNet (R50)} &
  \textbf{Unet ++ (R50)} &
  \textbf{Unet (Eff. Net)} &
  \textbf{UNet ++ (Eff. Net)} &
  \textbf{TransUNet} &
  \textbf{MedSAM} \\ \midrule
\multirow{4}{*}{\rotatebox{90}{\shortstack{Same-\\Dom.}}} & \textbf{BMC} &
  76.77 $\pm$ 0.47 &
  75.99 $\pm$ 1.21 &
  77.86 $\pm$ 1.12 &
  77.63 $\pm$ 1.00 &
  65.61 $\pm$ 2.72 &
  91.38 $\pm$ 0.69 \\
& \textbf{RUNMC} &
  77.92 $\pm$ 1.39 &
  78.62 $\pm$ 0.24 &
  79.33 $\pm$ 0.92 &
  79.79 $\pm$ 0.36 &
  66.17 $\pm$ 1.64 &
  83.75 $\pm$ 0.87 \\
& \textbf{BIDMC} &
  71.44  $\pm$ 1.28 &
  73.12 $\pm$ 0.41 &
  75.55 $\pm$ 0.66 &
  74.42 $\pm$ 0.45 &
  49.30 $\pm$ 8.22 &
  85.04 $\pm$ 1.80 \\
& \textbf{HK} &
  71.34 $\pm$ 0.85 &
  70.76 $\pm$ 1.48 &
  70.45 $\pm$ 1.43 &
  69.69 $\pm$ 3.50 &
  44.58 $\pm$ 4.49 &
  82.36 $\pm$ 1.78 \\ \midrule
\multirow{12}{*}{\rotatebox{90}{Cross-Dom. $S \rightarrow T$}} & \multirow{2}{*}{\textbf{BMC $\rightarrow$ RUNMC}} &
  63.37 $\pm$ 1.85 & 62.74 $\pm$ 2.67 & 58.29 $\pm$ 5.09 & 56.34 $\pm$ 5.43 & 21.51 $\pm$ 2.28 & 85.23 $\pm$ 1.19
  \\ & & ($\downarrow$ \textbf{13.4})& 
  ($\downarrow$ \textbf{13.3})&
  ($\downarrow$ \textbf{19.6})&
  ($\downarrow$ \textbf{21.3})&
  ($\downarrow$ \textbf{44.1})&
  ($\downarrow$ \textbf{6.1})
  \\
& \multirow{2}{*}{\textbf{BMC $\rightarrow$ BIDMC}} &
  51.22 $\pm$ 16.52  & 55.18 $\pm$ 3.78 & 46.35 $\pm$ 9.48 & 43.21 $\pm$ 7.95 & 4.67 $\pm$ 0.47 & 84.73 $\pm$ 1.05  
  \\ & & ($\downarrow$ \textbf{25.5})&
   ($\downarrow$ \textbf{20.8})&
   ($\downarrow$ \textbf{31.5})&
   ($\downarrow$ \textbf{34.4})&
   ($\downarrow$ \textbf{60.9})&
   ($\downarrow$ \textbf{6.6})
  \\
& \multirow{2}{*}{\textbf{BMC $\rightarrow$ HK}} &
  56.11 $\pm$ 4.93 & 55.34 $\pm$ 6.01 & 34.61 $\pm$ 6.66 & 36.11 $\pm$ 5.29 & 13.11  $\pm$ 4.66 & 83.40 $\pm$ 0.44 
  \\ & & ($\downarrow$ \textbf{20.7})&
  ($\downarrow$ \textbf{20.7})&
    ($\downarrow$ \textbf{43.3})&
    ($\downarrow$ \textbf{41.5})&
   ($\downarrow$ \textbf{52.5})&
    ($\downarrow$ \textbf{8.0})
   \\ \cmidrule{2-8}
& \multirow{2}{*}{\textbf{BIDMC $\rightarrow$ BMC}} &
   27.07 $\pm$ 1.67 & 24.86 $\pm$ 9.09 & 45.33 $\pm$ 6.39 & 42.41 $\pm$ 7.27 & 6.40 $\pm$ 1.33 & 90.77 $\pm$ 0.69
   \\ & & ($\downarrow$ \textbf{44.4})&
    ($\downarrow$ \textbf{48.3})&
   ($\downarrow$ \textbf{30.2})&
    ($\downarrow$ \textbf{32.0})&
   ($\downarrow$ \textbf{42.9})&
    ($\uparrow$ \textbf{5.7})
   \\
& \multirow{2}{*}{\textbf{BIDMC $\rightarrow$ RUNMC}} &
  4.41 $\pm$ 0.69 & 6.37 $\pm$ 1.41 & 20.81 $\pm$ 9.45 & 18.12 $\pm$ 11.04 & 4.19 $\pm$ 0.54 & 81.27 $\pm$ 1.05 
  \\ & & ($\downarrow$ \textbf{67.0})&
   ($\downarrow$ \textbf{66.8})&
   ($\downarrow$ \textbf{54.7})&
   ($\downarrow$ \textbf{56.3})&
   ($\downarrow$ \textbf{45.1})&
    ($\downarrow$ \textbf{3.8})
   \\
& \multirow{2}{*}{\textbf{BIDMC $\rightarrow$ HK}} &
  52.83 $\pm$ 2.62 & 50.16 $\pm$ 2.09 & 47.59 $\pm$ 1.52 & 48.23 $\pm$ 4.30 & 24.64 $\pm$ 4.94 & 81.19 $\pm$ 0.92 
  \\& & ($\downarrow$ \textbf{18.6})&
   ($\downarrow$ \textbf{23.0})&
   ($\downarrow$ \textbf{28.0})&
   ($\downarrow$ \textbf{26.2})&
   ($\downarrow$ \textbf{24.7})&
   ($\downarrow$ \textbf{3.9})
  \\ \bottomrule
\end{tabular}}
\label{tab:prostate}
\end{table}
\section{Experiment Results}
\vspace{-0.1in}
\subsection{Datasets and Implementations}
We experiment on three segmentation tasks, including Diabetic Retinopathy \textit{(DR) grading-related lesion segmentation} in 2D fundus images, \textit{liver structure segmentation} from 3D CT scans, and \textit{prostate segmentation} from 3D MRI data. Table \ref{tab:dataset_overview} in the Appendix provides details about each task, including information about the source and target domains, aiming to explore challenges related to domain shifts.
% Details of each task as well as the source and target domains to investigate the  domain shift challenges are included in Table \ref{tab:dataset_overview} Appendix. 
% We curate separate datasets for the source and target domains to investigate the challenge of domain shifts.  
% Table 1 provides an overview of the datasets used and the corresponding structures that were trained. 
We use the default training, testing split in each dataset if available; otherwise, we randomly select $80\%$ total samples for training and $20\%$ remaining for testing. On 3D segmentation problems, we reformulate them as independent 2D slice segmentation and merge results to a single 3D volume to compare with ground truths.

% We test two SAM's variations: \texttt{MedSAM} and \texttt{SAM-TransUNet} as described in Subsection\,\ref{subsec:prompt-end-2-end}. 
The large vision model SAM~\citep{kirillov2023segment} is bench-marked by utilizing the \texttt{MedSAM} method~\citep{huang2023segment} to fine-tune on a specific medical downstream task. It is important to emphasize that the SAM model was not previously pre-trained with extensive medical images, aligning with the approach taken by pre-trained ImageNet models
~\cite{deng2009imagenet}. We also compare SAM against popular supervised methods such as \texttt{TransUNet} with ViT-16 backbone \cite{dosovitskiy2020image}, \texttt{U-Net} \cite{ronneberger2015u} and \texttt{U-Net++} \cite{zhou2018unet++} with ResNet50 (R50) \cite{he2016deep}, and Efficient-Net b0 (Eff.Net) \cite{tan2019efficientnet}. All weights are initialized from ImageNet \cite{deng2009imagenet}. Details for uncertainty network $\bm{\delta}(\{\bold{\hat{y}}, \bold{x}\};\gamma^{*})$ are presented in Appendix.


\subsection{Quantitative Performance on Cross-domain}
\label{subsec:cross-domain}
We report the performance of various model architectures on different medical modalities in Tables~\ref{tab:prostate}, \ref{tab:IDRiD_FGADR-perform}, and \ref{tab:FLARE_LITS}. Each model was initially trained and evaluated within the \textit{same domain}. Subsequently, training is conducted on the source domain $S$, and evaluation is performed on the target domain $T$, adhering to the 
\textit{cross-domain} $S \rightarrow T$ setting.
% It is then trained on source domain $S$ and tested on target domain $T$ under the cross-domain $S \rightarrow T$ setting. 
Across all scenarios, \texttt{MedSAM} consistently demonstrates superior in- and out-of-domain performances, significantly surpassing other models. This noteworthy achievement highlights \texttt{MedSAM}'s robust OOD generalization capabilities, proven effective across both balanced (prostate modality) and imbalanced (diabetic retinopathy lesion and liver modalities) datasets. Further detailed analysis is available in the Appendix. To our latest knowledge, we first examine the robustness of the SAM model under domain shift in medical segmentation tasks, given models are fine-tuned on medical source domains~\citep{mazurowski2023segment,qiao2023robustness}.

\begin{table}[t!]
\centering
% \begin{wraptable}{L}{11cm}
\setlength{\tabcolsep}{8pt}
\renewcommand{\arraystretch}{0.4}
\captionsetup{%
   justification=justified,
%  singlelinecheck=off
}
\caption{Performance comparison in domain-shift of DR lesion segmentation. Results are reported in average 2D Dice score with three training times. Arrows $\uparrow$ and $\downarrow$ indicate the increase/drop performance.}
\vspace{2mm}
\label{tab:IDRiD_FGADR-perform}
\resizebox{0.9\columnwidth}{!}{
\begin{tabular}{l|cc|cc}
\toprule
& \multicolumn{2}{c|}{\textbf{Same-Domain}} & \multicolumn{2}{c}{\textbf{Cross-Domain $S \rightarrow T$}} \\ \cmidrule{2-5}
 & \textbf{IDRID} & \textbf{FGADR} & \textbf{IDRID $\rightarrow$ FGADR} & \textbf{FGADR $\rightarrow$ IDRID} \\ \midrule
\textbf{UNet (R50)} & 35.72 $\pm$ 1.35 & 49.46 $\pm$ 1.07 & 13.51 $\pm$ 5.56 \,($\downarrow$\,\textbf{22.21}) 
%($\downarrow$\,\textbf{22.2}) 
& 31.76 $\pm$ 1.61 \,($\downarrow$\,\textbf{17.7})\\
\textbf{Unet ++ (R50)} & 32.36 $\pm$ 3.70 & 48.73 $\pm$ 1.42 & 10.42 $\pm$ 2.19\,\,
($\downarrow$\,\textbf{21.9}) 
%($\downarrow$\,\textbf{38.31}) 
& 30.60 $\pm$ 1.31 ($\downarrow$\,\textbf{18.1}) \\
\textbf{Unet (Eff.Net)} & \multicolumn{1}{c}{34.18 $\pm$ 2.00} & {48.78 $\pm$ 0.59} & {12.72 $\pm$ 1.24}
($\downarrow$\,\textbf{21.5}) 
%($\downarrow$\,\textbf{36.06})
& \multicolumn{1}{l}{33.61 $\pm$ 1.29\,($\downarrow$\,\textbf{15.2})} \\
\textbf{UNet ++ (Eff.Net)} & \multicolumn{1}{l}{36.70 $\pm$ 1.61} & {49.88 $\pm$ 0.85} & \multicolumn{1}{l}{\,\,20.51 $\pm$ 1.41\,
($\downarrow$\,\textbf{16.2})} 
%($\downarrow$\,\textbf{29.37})}
& \multicolumn{1}{l}{32.49 $\pm$ 2.08\, ($\downarrow$\,\textbf{17.4})} \\ 
\textbf{TransUNet} & \multicolumn{1}{l}{15.28 $\pm$ 1.62} & {46.59 $\pm$ 0.80} & \multicolumn{1}{l}{\ 11.58 $\pm$ 11.04\ ($\downarrow$\,\textbf{3.7})} & \multicolumn{1}{l}{22.52 $\pm$ 2.39\ \ ($\downarrow$\,\textbf{24.1})} \\ \midrule
\textbf{MedSAM} & 37.76 $\pm$ 1.37 & 58.49 $\pm$ 0.29 & 44.63 $\pm$ 1.42 \,($\uparrow\textbf{\underline{6.9}}$) & 39.24 $\pm$ 0.62 \,($\downarrow$ \textbf{19.25}) \\  \bottomrule
\end{tabular}}
\end{table}
% \vspace{-0.3in}
\begin{table}[t]
\centering
\caption{Performance comparison in domain-shift of liver segmentation. Results are computed by average 2D Dice score in three training times. Arrows $\uparrow$ and $\downarrow$ indicate the increase/drop performance.}
\vspace{2mm}
\label{tab:FLARE_LITS}
\resizebox{0.9\columnwidth}{!}{
\begin{tabular}{l|cc|cc}
\toprule
& \multicolumn{2}{c|}{\textbf{Same-Domain}} & \multicolumn{2}{c}{\textbf{Cross-Domain} $S \rightarrow T$} \\ \cmidrule{2-5}
 &
  \textbf{FLARE} &
  \textbf{LiTS} &
  \textbf{FLARE $\rightarrow$ LiTS} &
  \textbf{LiTS $\rightarrow$ FLARE} \\ \midrule
\textbf{UNet (R50)}       & 94.35 $\pm$ 1.16 & 95.69 $\pm$ 0.09  & 82.28 $\pm$ 4.41\,($\downarrow$\,\textbf{12.1}) &  95.57 $\pm$ 0.39\,($\downarrow$\,\textbf{0.1})\\
\textbf{Unet ++ (R50)}    & 96.08 $\pm$ 0.40 & 95.84 $\pm$ 0.12 & 72.51 $\pm$ 5.97 \,($\downarrow$\,\textbf{23.6})  & 95.41 $\pm$ 0.45\,($\downarrow$\,\textbf{0.4})  \\
\textbf{Unet (Eff. Net)}    &  95.11 $\pm$ 0.18 & 95.86 $\pm$ 0.23 & 68.91 $\pm$ 7.06\,($\downarrow$\,\textbf{26.2}) &  95.04 $\pm$ 0.63\,($\downarrow$\,\textbf{0.8})\\
\textbf{UNet ++ (Eff. Net)} & 95.23 $\pm$ 0.98 & 95.57 $\pm$ 0.40 & 67.32 $\pm$ 3.27\,($\downarrow$\,\textbf{27.91})  &  95.0 $\pm$ 0.95\,($\downarrow$\,\textbf{0.6})\\ 
\textbf{TransUNet}              & 92.01 $\pm$ 0.78 & 92.22 $\pm$ 3.22 & 61.69 $\pm$ 1.17\,($\downarrow$\,\textbf{30.3}) & 93.2 $\pm$ 1.68\,($\uparrow$\,\textbf{1.0}) \\ \midrule
\textbf{MedSAM} &
  92.47 $\pm$ 0.02 & 97.80 $\pm$ 0.01
   & 97.53 $\pm$ 0.03 \,($\uparrow$\,\textbf{\underline{5.1}})
   & 92.17 $\pm$ 0.24 \,($\downarrow$\,\textbf{5.6})
   \\  \bottomrule
\end{tabular}}
\end{table}

\subsection{Estimating Out-of-Distribution Performance via Uncertainty Perspective}
We now investigate different indicators to characterize the models' OOD properties. We consider both popular indicators used in the natural image domains, such as the ID performance \cite{miller2021accuracy}, the agreement~\cite{baek2022agreement}, and our proposed Bayesian uncertainty. 

\begin{figure}
    \centering
    \includegraphics[width=\textwidth]{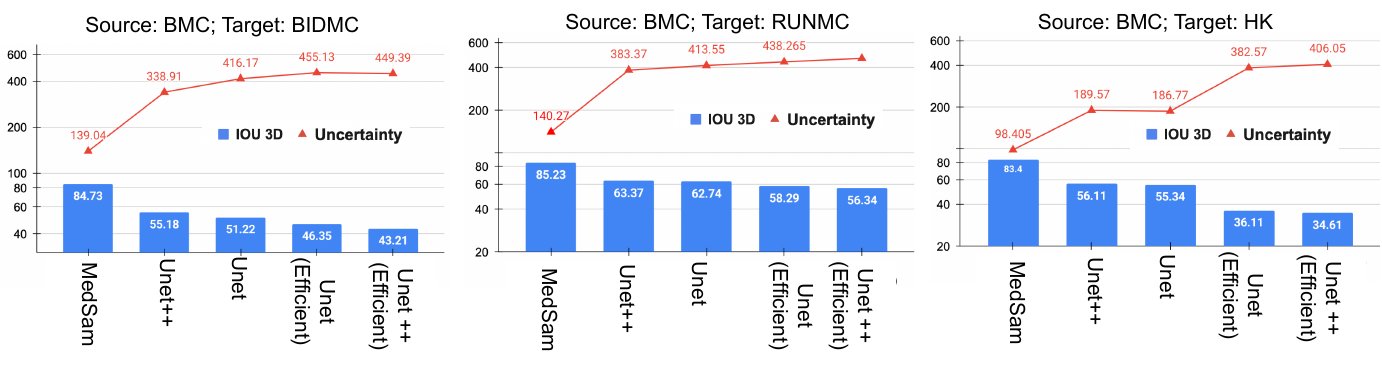}
    \caption{Visualization of uncertainty and OOD performance on three cross-domain experiments in prostate segmentation. Uncertainty exhibits a strong correlation with OOD performance, indicating that higher uncertainty values tend to correspond to lower OOD performance.} 
    \vspace{-2mm}
    \label{ood-uncertainty}
\end{figure}

Figure \ref{ood-uncertainty} illustrates a strong correlation between uncertainty and OOD performance in three settings of prostate segmentation (Table~\ref{tab:prostate}), underscoring that elevated uncertainty values tend to align with diminished OOD performance. Additionally, when comparing different models on the same dataset, the OOD performance gaps between these models are found to be correlated with the gaps in their uncertainties. However, it is important to highlight that uncertainty only approximated the true error in this context. As a result, even when two models exhibit similar OOD performances, there may still be some subtle differences in their uncertainties, reflecting nuanced variations in their predictive capabilities. This phenomenon presents an interesting future research direction for accurately estimating a model OOD performance.
%This underscores the importance of considering uncertainty as a complementary metric to assess model reliability and robustness in OOD scenarios.

\begin{figure}
    \centering
    \includegraphics[width=\textwidth]{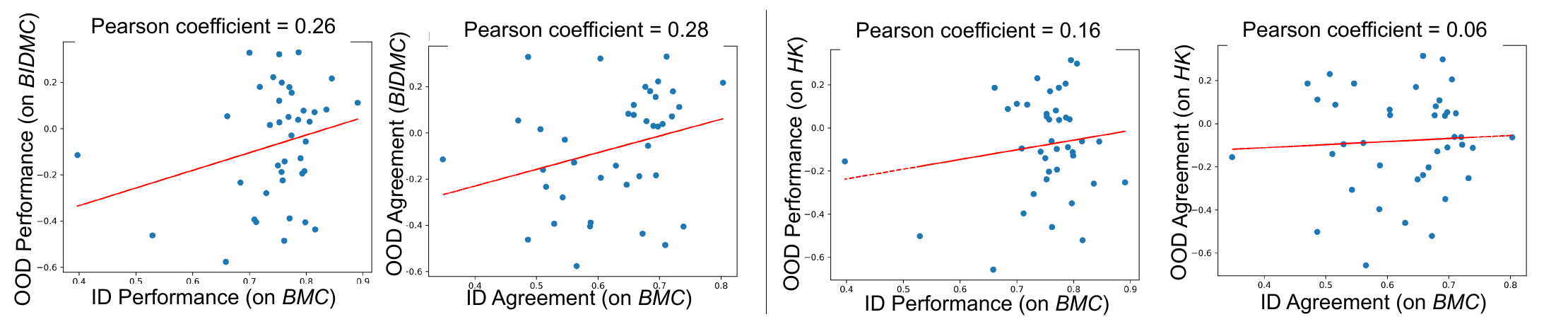}
    \caption{OOD performance/agreement versus ID performance/agreement on BMC and HK datasets (left) and on BMC and BIDMC datasets (right).} 
    \vspace{-5mm}
   \label{fig:id-ood-linear}
\end{figure}

\paragraph{\textbf{Comparing with Other Correlations}:}
Figure~\ref{fig:id-ood-linear} reports the correlations between the OOD performance and the ID performance or the ID agreement when using BMC as the source domain and BIDMC as the target domain. By varying the training configurations, such as learning rates, epochs, etc., we plot the OOD performance against the ID indicator for each model. Afterward, we compute Pearson correlation coefficients between the OOD performance and ID indicators. This analysis aims to substantiate whether linear correlations, as proposed in previous studies \citep{miller2021accuracy,baek2022agreement}, are indeed observed. Nevertheless, none of these indicators proves effective in adequately characterizing the OOD performance, as they only yield relatively low Pearson correlation coefficients. Consequently, we can infer that the currently employed natural image indicators are not well-suited for our specific medical image segmentation task.

% \subsubsection{Comparing with Uncertainty Baselines}
% \begin{table}[t]
\begin{wraptable}{L}{8cm}
 %\RaggedRight
\setlength{\tabcolsep}{6pt}
\renewcommand{\arraystretch}{0.5}
\captionsetup{%
   justification=justified,
   % labelfont=bf,
%  singlelinecheck=off
}
\caption{Comparison among uncertainty approaches \\ measured by mean squared error. Smaller is better.}
\label{tab:uncertainty_compare}
% \vspace{2mm}
\resizebox{0.53\columnwidth}{!}{
\begin{tabular}{ @{\hspace{-0pt}}l@{\hspace{8pt}}c@{\hspace{8pt}}c@{\hspace{10pt}}c@{\hspace{8pt}}c}
\toprule
\textbf{Method}         & \textbf{BMC} $\rightarrow$ \textbf{RUNMC} & \textbf{BMC} $\rightarrow$ \textbf{BIDMC} & \textbf{BMC} $\rightarrow$ \textbf{HK} \\
\midrule
\textbf{Our}                   &  \textbf{0.012}    & \textbf{0.011} & \textbf{0.018} \\
BayesCap                 &  0.015  &  0.017 & 0.021 \\
TTDAp              & 0.247  &  0.326 & 0.387 \\
TTDAc             & 0.212   & 0.248 & 0.291  \\
DropOut           &   0.197 &  0.227 & 0.271 \\
\bottomrule
\end{tabular}}
\end{wraptable}
    % \end{table}
\paragraph{\textbf{Comparing with Uncertainty Baselines}:}
We furthermore compare our approach against two different baseline groups, including Bayesian identity mapping \cite{upadhyay2022bayescap} and perturbing the original images to gauge the uncertain level for the SAM model.
% which are learning the Bayesian identity mapping \cite{upadhyay2022bayescap} and perturbing the original images. 
The former group, known as BayesCap \cite{upadhyay2022bayescap}, calibrates the uncertainties of a pre-trained model without the need to train itself on a large-scale dataset. The latter is based on \textit{test-time modification} techniques~\citep{ayhan2022test,wang2022test,wang2019aleatoric}. Herein, we employ two types of modifications: test-time data augmentation (TTDA) \cite{ayhan2022test,wang2022test} and dropout \cite{srivastava2014dropout} before the final prediction~\cite{kendall2015bayesian,gal2016dropout}. Two classes of data augmentation are adapted for baselines, including pixel-wise noise (TTDAp) and color jittering (TTDAc). Table \ref{tab:uncertainty_compare} illustrates various methods' performance in three cross-domain prostate segmentation settings. The results are presented as mean square errors between the uncertainty values, estimated using the Otsu thresholding algorithm, and the actual errors of the model. Notably, our approach consistently outperforms others, attaching the top records across all three Out-of-Domain (OOD) settings.

\section{Conclusion}
\vspace{-0.1in}
In this study, we first conducted a systematic investigation into the performance of the foundation model SAM under domain-shift segmentation tasks. Our findings reveal that this foundation model demonstrates better generalization capabilities than other methods initialized from pre-trained ImageNet. Additionally, we show the correlation between model uncertainties and their out-of-distribution performance; thereby,
lower uncertainty predictions tend to reflect a higher performance in OOD. To achieve this, we constructed a post-hoc estimation approach tailored for pre-trained deterministic models. These models have demonstrated the ability to consistently generate well-calibrated uncertainty estimates across various segmentation scenarios, proving their utility in aiding experts during the decision-making process.
% the post-hoc uncertainty estimation method for pre-trained deterministic models that are shown to produce consistently well-calibrated uncertainty estimates across segmentation settings and be useful in assisting experts in decision-making. 
One limitation of our algorithm is that the uncertainty can only approximate the true errors; therefore, it may not be suitable to compare models whose performances have small margins. 
% the estimated domain gaps using the maximum mean discrepancy (MMD) distance and the observed performance variations of SAM models. We discovered that MMD is particularly relevant to performance changes in several cases. 
In future work, we plan to improve this issue as well as explore the proposed method in a broader range of foundation methods with diverse settings to gain better insights into the behaviors of the proposed algorithms.

\section*{Acknowledgements}
This research has been supported by the pAItient project (BMG, 2520DAT0P2), Ophthalmo-AI project (BMBF, 16SV8639), and  Endowed Chair of Applied Artificial Intelligence, Oldenburg University, Germany. Tan Ngoc Pham would like to thank the Vingroup Innovation Foundation (VINIF) for the Master's training scholarship program. The authors thank the International Max Planck Research School for Intelligent Systems (IMPRS-IS) for supporting Duy M. H. Nguyen and Vinh Tong. Mathias Niepert acknowledges funding by Deutsche Forschungsgemeinschaft (DFG, German Research Foundation) under Germany’s Excellence Strategy - EXC and support by the Stuttgart Center for Simulation Science (SimTech).

\bibliography{references}

\clearpage
\appendix
\begin{center}
    {\bf {\Large Supplementary Material for ``On the Out of Distribution Robustness of Foundation Models in Medical Image Segmentation"} }
\end{center}

\section{Background}
\subsection{SAM Architecture}
\label{subsec:sam-architecture}
SAM \cite{kirillov2023segment} was introduced by Meta AI to improve segmentation performance across a wide range of images. SAM utilizes a transformer-based architecture \cite{vaswani2017attention}, which has shown impressive achievement in both natural language processing \cite{brown2020language} and computer vision \cite{dosovitskiy2020image}. 
In general, SAM is a Vision Transformer (ViT)-based model, which consists of three components: an image encoder, a prompt encoder, and a mask decoder. The image encoder is based on ViT \cite{dosovitskiy2020image}, which is pre-trained with masked auto-encoder (MAE) \cite{he2022masked}. Two sets of prompts can be considered, including sparse prompts (points, boxes, text) and dense prompts (masks). To represent the first two prompts, points, and boxes, SAM employs positional encoding \cite{tancik2020fourier} combined with learned embeddings. 
% Particularly, a point is encoded by two learnable tokens indicating foreground and background, and a bounding box is encoded by the point encodings of its top-left corner and bottom-right corner.
The text prompt is encoded by the pre-trained text-encoder CLIP \cite{radford2021learning}. The mask prompt has the same spatial resolution as the input image, and it is encoded using convolutions and summed element-wise with the image embedding. The mask decoder uses a lightweight network consisting of two transformer layers and a dynamic mask prediction head with an Intersection-over-Union (IoU) score regression head. SAM is trained in a supervised learning manner on a large-scale SA-1B dataset with over 1 billion masks from 11 million natural images using a linear combination of Dice loss \cite{milletari2016v} and Focal loss \cite{lin2017focal}.

\subsection{Implementation Details}
% We test two variations of SAM as described in Subsection\,\ref{subsec:prompt-end-2-end}, denoted as \texttt{MedSAM} and \texttt{SAM-TransUNet}. 
We use \texttt{MedSAM} to fine-tune the SAM model on a specific medical downstream task.
The Adam optimizer is utilized for training networks with a combination of Dice loss and Cross-entropy. The learning rates for problems are selected from a set of $\{1e\mathrm{-}4, 3e\mathrm{-}4, 5e\mathrm{-}4, 1e\mathrm{-}5\}$ depend on validation performance.  

% On 3D segmentation problems, we reformulate them as independent 2D slice segmentation and merge results to a single 3D volume to compare with ground truths. We additionally compare SAM models against popular supervised methods such as \texttt{TransUNet} with ViT-16 backbone \cite{dosovitskiy2020image}, \texttt{U-Net} \cite{ronneberger2015u} and \texttt{U-Net++} \cite{zhou2018unet++} with ResNet50 (R50) \cite{he2016deep}, and Efficient-Net b0 (Eff.Net) \cite{tan2019efficientnet}. All weights are pre-trained on ImageNet. 

To gauge the extent of uncertainty arising from the foundational models, we utilize a deep architecture named $\bm{\delta}(\{\bold{\hat{y}}, \bold{x}\};\gamma^{*})$, inspired by the ResNet family \cite{he2016deep} with 18 convolutional layers. In our experiment, we use Eqs. \eqref{eq:alpha_beta_ggd_1},\eqref{eq:alpha_beta_ggd_2} for training with $\bm{\beta}_t$ fixed as $\textbf{2}_{\textbf{m}}$. The initial input for the model $\bm{\delta}$ is a joint feature vector constructed by concatenating the original image $\bold{x}$ with the prediction mask $\hat{\bold{y}}$ after they have passed through two different convolutional layers. Additionally, the output of the model $\bm{\delta}$ is also combined with the aforementioned joint feature vector. These combined features then undergo processing through distinct convolutional operators, ultimately producing parameters representing uncertainty distributions
% two final masks: $\bold{\mu}$ and $\bold{\sigma}$ 
as described in Eq.(\ref{eq:uncertainty_model}). In the end, the uncertainty model $\bm{\delta}$ is trained with Adam optimizer \cite{kingma2014adam} with the learning rate of $1e^{-4}$ in 50 epochs and cosine annealing strategy \cite{loshchilov2016sgdr} for the learning rate warm-up.

\section{Datasets}
Table \ref{tab:dataset_overview} overviews used dataset in our experiment along with the modality and image types. Each task has at least two datasets, one for a source domain and the remaining for a target domain. 
%\begin{wraptable}{l}{9cm}
\begin{table}[!ht]
\vspace{-0.2in}
\caption{Overview datasets used in our experiment.}
\vspace{1mm}
\label{tab:dataset_overview}
\centering
\resizebox{.7\columnwidth}{!}{
\begin{tabular}{c|c|c|c|c}
\toprule
\textbf{Task} & \textbf{Objects} & \textbf{Datasets} & \textbf{Modality} & \textbf{\# Images} \\ \midrule
\multirow{2}{*}{\shortstack{DR Lesion \\Segmentation}} & \multirow{2}{*}{HE, SE, EX, MA} & FGADR \cite{zhou2020benchmark} & 2D Fundus & 1842 \\
     &        & IDRiD \cite{porwal2018indian}  & 2D Fundus   & 81   \\ \midrule
\multirow{2}{*}{\shortstack{Liver\\Segmentation}}     & \multirow{2}{*}{Liver}          & FLARE \cite{simpson2019large} & 3D CT  & 50   \\
     &        & LiTS  \cite{bilic2023liver}  & 3D CT    & 130  \\ \midrule
\multirow{4}{*}{\shortstack{Prostate \\Segmentation}}  & \multirow{4}{*}{Prostate}       & BMC \cite{liu2020ms}  & 3D MRI & 30   \\
     &        & BIDMC \cite{liu2020ms}  & 3D MRI   & 12   \\
     &        & RUNMC \cite{liu2020ms}  & 3D MRI   & 30   \\
     &        & HK  \cite{liu2020ms}    & 3D MRI   & 12   \\ \bottomrule
\end{tabular}}\hspace{1.5em}
\end{table}
%\end{wraptable}
\section{Experiment Details \& Analyses}
\subsection{Further Analysis on Cross-domain Performance}
\label{sec:further_analysis}
\vspace{-0.1in}
The performance comparisons are presented in Tables \ref{tab:prostate}, \ref{tab:IDRiD_FGADR-perform}, and \ref{tab:FLARE_LITS} for three different medical modalities: Prostate, DR, and Liver, respectively. For each dataset, we report the performance on both Unet-based models (including Unet (R50), Unet++ (R50), Unet (Eff.Net), Unet++ (Eff.Net)), and SAM-based model (i.e., MedSAM). Each model was initially trained and tested within the same domain. It is then trained on source domain $S$ and tested on target domain $T$ under the cross-domain $S \rightarrow T$ settings. 

Table \ref{tab:prostate} demonstrates that MedSAM not only outperforms other models but also exhibits superior generalization with minimal performance gaps between the same domain and cross-domain. 
%While SAM-TransUnet does not achieve a mean Dice score as high as MedSAM, it still outperforms TransUnet alone, particularly in the cross-domain scenarios. The performance gaps between the same-domain and cross-domain settings obtained by TransUnet and SAM-TransUnet highlight the enhanced generalization capabilities of SAM-TransUnet through the utilization of SAM. 

Four datasets (BMC, BIDMC, RUNMC, and HK) on Prostate modality are well-balanced in terms of the number of images. Thus, the domain shift has been clearly observed, as shown in Table\,\ref{tab:prostate}. On the other hand, the DR Lesion datasets (FGADR and IDRiD) and Liver datasets (FLARE and LiTS) suffer from heavy class imbalance, as indicated in Table\,\ref{tab:dataset_overview}. Notably, the Liver datasets consist of 3D CT Scans processed slide-by-slide, with FLARE having 3,080 slides and LiTS having 25,660 slides. In Table \ref{tab:IDRiD_FGADR-perform}, MedSAM achieves state-of-the-art performance with small domain gaps. 
% , while SAM-TransUnet outperforms TransUnet with similar domain gaps. Our hypothesis is that the limited size of FGADR and IDRiD datasets hinders substantial improvement in model generalization through fine-tuning. Furthermore, MedSAM used bounding boxes-based prompts during the inference step; thus, it is more robust to noise from the surrounding context than SAM-TransUnet. 
Due to the large imbalance between LiTS (25,660 slides) and FLARE (3,080 slides), models trained on LiTS, whether Unet-based or SAM-based, demonstrate better generalization compared to those trained on FLARE, as demonstrated in Table \ref{tab:FLARE_LITS}, except MedSAM. While MedSAM shows a great improvement on small datasets such as FLARE (i.e., FLARE $\rightarrow$ LiTS), its generalization is dropped on large dataset LiTS. 

\subsection{Qualitative Results}

\begin{figure}[H]
    \centering
    \subfloat{
      \includegraphics[width=0.17\textwidth]{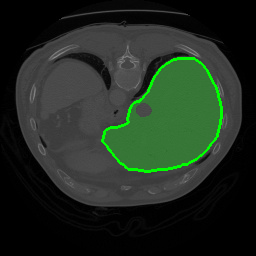}
      \includegraphics[width=0.17\textwidth]{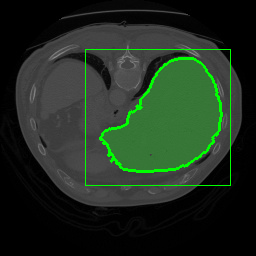}
    }
    \subfloat{
      \includegraphics[width=0.17\textwidth]{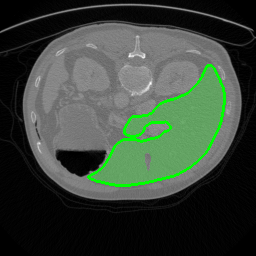}
      \includegraphics[width=0.17\textwidth]{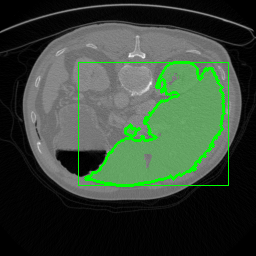}
    } 
    \caption{A visual demonstration on MedSAM's performance in the same- and cross-domain. The two on the left depict the true mask and MedSAM's prediction mask in the same domain (FLARE). The other two display the true mask and MedSAM's prediction mask in an out-of-domain scenario (FLARE $\rightarrow$ LiTs). Best views in color with zoom.}
    \label{fig:Medsam_demo}
\end{figure}

\begin{figure}[H]
    \centering
    \includegraphics[width=0.8\textwidth]{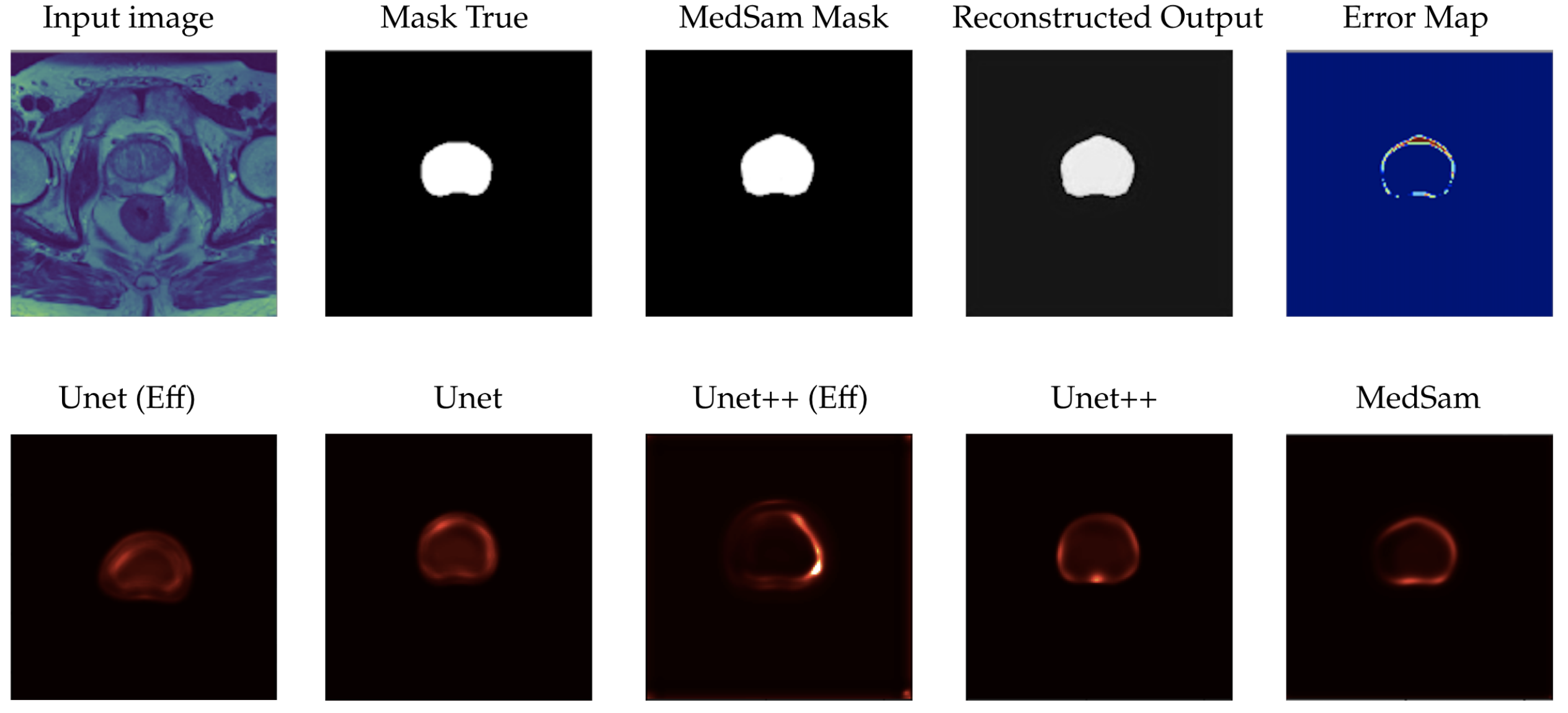}
    \caption{A qualitative comparison among various foundation models. Respectively, from left to right, the top row demonstrates the original image, the true mask for that image, MedSAM's prediction mask, the reconstructed mask from the uncertainty model, and the error map between the true mask and MedSAM's prediction mask. The bottom row ranks the uncertain levels of different foundation models from the most uncertainties (leftmost) to the fewest uncertainties (rightmost), whereas the fewer, the better. Following this, the prediction mask derived from MedSAM has the fewest uncertainties.} 
    \label{fig:uncertainty}
\end{figure}

In Figure \ref{fig:Medsam_demo}, we present a selection of predictions generated by the MedSAM model in both same and cross-domain settings, showcasing the segmentation results for lung, prostate, and soft exudates. Given the prompt-based nature of MedSam, we simulate bounding boxes-based prompts as \cite{ma2023segment}. Our observations indicate that MedSAM is capable of producing satisfactory masks for prostate structures and DR lesions (shown in the second and bottom rows) despite the challenging conditions of small object sizes and domain shifts. However, when applied to the LiTS dataset, MedSAM tends to exhibit tendencies of over-segmenting boundaries or missing structures at the bends of objects.

We offer illustrative examples of uncertainty estimations generated by our algorithm, showcased in Figure \ref{fig:uncertainty}, across various trained models. It can be seen the MedSam model has the highest relevant uncertainty regions compared to the Error map among models.
% In Figure \ref{fig:uncertainty}, we present illustrative examples of uncertainty estimations generated by our algorithm across various trained models. Notably, the MedSam model exhibits the most pronounced uncertainty regions, surpassing those depicted in the Error map.
\end{document}